\documentclass[%
 reprint,longbibliography,preprintnumbers,
nofootinbib,
 amsmath,amssymb,
 aps
pre
]{revtex4-1}
\pdfoutput=1
\usepackage{graphicx}
\usepackage[utf8]{inputenc}
\usepackage{flushend}
\usepackage{dcolumn}
\usepackage{bm}
\usepackage{balance}

\usepackage[normalem]{ulem}

\usepackage[colorlinks = true,
            linkcolor = blue,
            urlcolor  = blue,
            citecolor =green,
            anchorcolor = blue]{hyperref}
\usepackage{verbatim}
\usepackage{color,ulem}
\usepackage[english]{babel}

\usepackage[utf8]{inputenc}
\input Starburst.fd
\newcommand*\initfamily{\usefont{U}{Starburst}{xl}{n}}\initfamily

\newcommand{\beq}{\begin{eqnarray}}
\newcommand{\eeq}{\end{eqnarray}}
\usepackage{amsmath}
\usepackage{tikz}
\usetikzlibrary{decorations.pathmorphing}
\usetikzlibrary{shapes.misc}
\tikzset{cross/.style={cross out, draw=black, minimum size=8*(#1-\pgflinewidth), inner sep=0pt, outer sep=0pt},
cross/.default={1pt}}
\usetikzlibrary{patterns,math}
\begin{document}

\title{\Large  Revealing the supercritical dynamics of dusty plasmas and their liquid-like to gas-like dynamical crossover}

\author{Dong Huang$^{1}$}
\author{Matteo Baggioli$^{2}$} 
\email{b.matteo@sjtu.edu.cn}
\author{Shaoyu Lu$^{1}$}
\author{Zhuang Ma$^{1}$}
\author{Yan Feng$^{1}$}
\email{fengyan@suda.edu.cn}

\vspace{1cm}

\affiliation{$^{1}$Institute of Plasma Physics and Technology, School of Physical
Science and Technology, Jiangsu Key Laboratory of Thin Films, Soochow University, Suzhou 215006, China}
\affiliation{$^{2}$Wilczek Quantum Center, School of Physics and Astronomy, Shanghai Jiao Tong University, Shanghai 200240, China \& Shanghai Research Center for Quantum Sciences, Shanghai 201315, China}

\begin{abstract}
Dusty plasmas represent a powerful playground to study the collective dynamics of strongly coupled systems with important interdisciplinary connections to condensed matter physics. Due to the pure Yukawa repulsive interaction between dust particles, dusty plasmas do not display a traditional liquid-vapor phase transition, perfectly matching the definition of a supercritical fluid. Using molecular dynamics simulations, we verify the supercritical nature of dusty plasmas and reveal the existence of a dynamical liquid-like to gas-like crossover which perfectly matches the salient features of the Frenkel line in classical supercritical fluids. We present several diagnostics to locate this dynamical crossover spanning from local atomic connectivity, shear relaxation dynamics, velocity autocorrelation function, heat capacity, and various transport properties. All these different criteria well agree with each other and are able to successfully locate the Frenkel line in both 2D and 3D dusty plasmas. In addition, we propose the unity ratio of the instantaneous transverse sound speed $C_T$ to the average particle speed $\bar{v}_{p}$, i.e., $C_T / \bar{v}_{p} = 1$, as a new diagnostic to identify this dynamical crossover. Finally, we observe an emergent degree of universality in the collective dynamics and transport properties of dusty plasmas as a function of the screening parameter and dimensionality of the system. Intriguingly, the temperature of the dynamical transition is independent of the dimensionality and it is found to be always $20$ times of the corresponding melting point. Our results open a new path for the study of single particle and collective dynamics in plasmas and their interrelation with supercritical fluids in general.
\end{abstract}

\maketitle

\section{Introduction}
A supercritical fluid~\cite{Cowan:1988,Eckert:1996,Kessel:2005} typically refers to a condensed state of matter in which the traditional liquid and gas phases cannot be separated anymore by a sharp first-order phase transition. For various substances, the supercritical fluid state can be achieved when the temperature and the pressure are above the corresponding critical point~\cite{Cowan:1988,Eckert:1996,Kessel:2005}. Supercritical fluids have been intensively investigated due to their widely applications in the nuclear waste, petrochemical, food, and pharmaceutical industries~\cite{Carleson:1996, McHardy:1998, Kiran:2000}. Although there is no traditional liquid or gas state within the supercritical regime, recently, several studies suggest that a liquid-like to gas-like dynamical transition can be identified in supercritical fluids using either the Frenkel line~\cite{Brazhkin:2012,Brazhkin:2013,Cockrell:2021,Yang:2015,Bryk:2017,Bolmatov:2013,Lee:2021}, the Widom line~\cite{Simeoni:2010}, or the Fisher-Widom line~\cite{Fisher:1969}.\\

In the microscopic description of liquids proposed by Frenkel~\cite{Frenkel:1946}, atomic particle motion is a combination of quasiharmonic vibrations around potential minima and thermally induced jumps from an equilibrium position to a new one. These hopping processes give the ability to flow in liquids and happen at an average time $\tau$, also termed the liquid relaxation time~\cite{Brazhkin:2012,Frenkel:1946,Brazhkin:2013}. This view, deeply inspired by the phenomenological ideas of Maxwell \cite{doi:10.1098/rstl.1867.0004}, implies that liquids behave effectively as solids for time scales shorter than $\tau$, or equivalently, for frequencies larger than the Frenkel frequency $\omega_F=2\pi/\tau$. Following this picture inspired by solid state theory, the minimal period of rigid-like vibrations is given by the Debye time $\tau_D$, which is around $0.1-1$ ps in classical liquids. When $\tau > \tau_D$, particles mainly vibrate at their equilibrium positions and hop rarely, so that the typical liquid behavior is exhibited, often termed as the ``rigid liquid'' state~\cite{Brazhkin:2012,Brazhkin:2013}. In that regime, liquids are expected to support propagating shear waves at frequencies $\omega>\omega_F$. One can further derive a minimal cutoff wave-vector for their propagation \cite{BAGGIOLI20201}, which implies a maximum propagation length approximately equal to the sound speed times the relaxation time $\tau$.

Within this framework, the frequency of collective propagating shear waves in liquids, responsible for the emergent ''rigidity'', has to fall in the range $\omega_F<\omega<\omega_D$, with $\omega_D=2\pi/\tau_D$. The relaxation time $\tau$ decreases with the temperature. Physically, this is just reflected in a larger kinetic energy and therefore a stronger ability to rearrange. When $\tau < \tau_D$, the particles' hopping occurs more frequently than solid-like vibrations, collective shear waves disappear, and the ``nonrigid'' gas-like fluid state~\cite{Brazhkin:2012,Brazhkin:2013} is approached. Following this logic, in supercritical liquids, the concept of Frenkel line~\cite{Brazhkin:2012,Brazhkin:2013} has been introduced to discriminate the rigid liquid from the nonrigid gas-like fluid state and it has been formally defined by the condition $\tau \approx \tau_D$. In the past~\cite{Brazhkin:2012,Brazhkin:2013,Bolmatov:2013, Yang:2015, Cockrell:2021}, several diagnostics, including the specific heat $c_V$, the velocity autocorrelation function (VACF), and the mean squared displacement have been used to determine the condition for the Frenkel line in various physical systems, as described in detail later. Furthermore, when $\tau \rightarrow \tau_D$, i.e., at the onset of the dynamical crossover, the propagation length of shear waves approaches the minimum available value approximately given by the interatomic distance. Therefore, the Frenkel line can be also defined as the disappearance of collective shear waves in fluids \cite{BAGGIOLI20201} and it is related to universal minimal values for different transport coefficients such as the shear viscosity and the thermal diffusivity~\cite{doi:10.1126/sciadv.aba3747,PhysRevB.103.014311}.

\begin{figure}[h!]
	\centering
	\includegraphics[width=0.9\linewidth]{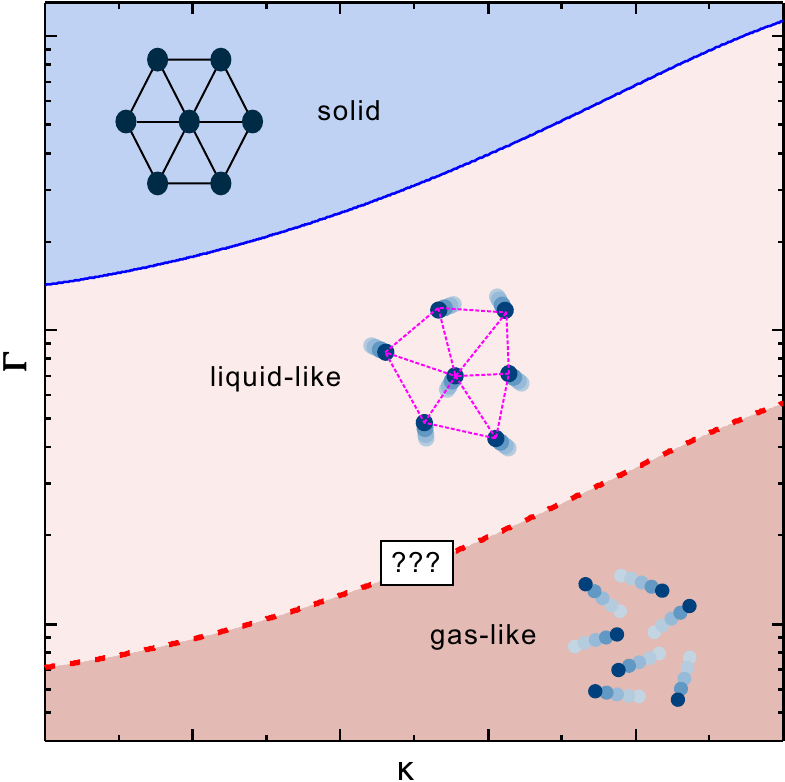}
	\caption{\label{fig1} A putative phase diagram for dusty plasmas as a function of the coupling parameter $\Gamma$ and the screening parameter $\kappa$. The solid blue line indicates the first-order solid-liquid phase transition. The dashed red line refers to a possible dynamical crossover in the supercritical regime which is the subject of this work.}
\end{figure}

Dusty plasmas~\cite{Thomas:1996, LI:1996, Melzer:1996, Merlino:2004, Kalman:2004, Morfill:2009, Piel:2010, Bonitz:2010}, also termed complex plasmas, are partially ionized gases containing micron-sized dust particles, and they represent an excellent model system where the motion of individual dust particles can be directly tracked. In the laboratory conditions, these dusts have a typical charge of $- 10^5 e$ in the steady state~\cite{Nosenko:2004, Feng:2010, Feng:2012}, interacting with each other through the Yukawa repulsion~\cite{Konopka:2000}, and leading to a much higher potential energy between neighboring dusts than their kinetic energy, i.e., these dusts are strongly coupled~\cite{Merlino:2004, Kalman:2004, Morfill:2009, Piel:2010, Bonitz:2010, Ichimaru:1982} (cf. classical liquids). During experiments, these dusts can form into either a single layer two-dimensional (2D) suspension or a three-dimensional (3D) suspension, i.e., 2D~\cite{Nunomura:2002,Nosenko:2004,Nosenko:2008,Hartmann:2010,Couedel:2010,Kahlert:2012,Hartmann:2014,Hartmann:2019,Feng:2010,Feng:2012,Du:2019,Gogia:2017} or 3D dusty plasma~\cite{Pramanik:2002,Tsai:2016,Hayashi:1999,Arp:2004,Bonitz:2006,Zuzic:2000,Morfill:2006,Thomas:2005}, exhibiting collective solid and liquid-like behaviors~\cite{Thomas:1996, LI:1996, Melzer:1996, Merlino:2004, Kalman:2004, Morfill:2009, Piel:2010, Bonitz:2010,Nunomura:2002,Nosenko:2004,Nosenko:2008,Hartmann:2010,Couedel:2010,Kahlert:2012, Hartmann:2014,Hartmann:2019,Feng:2010,Feng:2012,Hayashi:1999,Arp:2004,Bonitz:2006,Zuzic:2000,Morfill:2006,Pramanik:2002,Tsai:2016,Du:2019,Thomas:2005,Gogia:2017}, including the solid-liquid phase transition or melting~\cite{Melzer:1996, Nosenko:2004, Feng:2010}. Thus, dusty plasmas provide an incredibly powerful platform to explore collective dynamics of liquids and solids at the individual particle level~\cite{Nunomura:2002,Nosenko:2004,Nosenko:2008,Hartmann:2010,Couedel:2010,Kahlert:2012,Hartmann:2019,Hartmann:2014,Feng:2010,Feng:2012,Du:2019,Gogia:2017}. \\

In the past thirty years, the solid-liquid phase transition of dusty plasmas has been systematically investigated, in both experiments and simulations, as well as theories~\cite{Melzer:1996, Nosenko:2004, Feng:2010}. Commonalities between fluid dusty plasmas and classical liquids, rooted in their shared strongly coupled nature, have been also explored, creating a beneficial exchange of ideas and results. As a concrete example, the dynamics of shear waves and the existence of a critical wave-vector, typical of liquids \cite{Trachenko_2016,BAGGIOLI20201}, have been the subject of several theoretical and experimental studies in dusty plasmas \cite{PhysRevLett.85.2514,Goree:2012,PhysRevLett.97.115001}. However, so far, one important aspect of the collective dynamics of dusty plasmas remains elusive. In fact, due to the pure repulsive interaction between dust particles, there is no liquid-vapor phase transition~\cite{Hansen:1986} in dusty plasmas. Thus, a melted or fluid dusty plasma perfectly matches the definition of a supercritical fluid and should be thought as such. To the best of our knowledge, the supercritical nature of fluid dusty plasmas and the existence of a related liquid-like to gas-like dynamical crossover (Frenkel line) have been never disclosed before. In this work, we put forward this new interpretation and we further show that, as in classical liquids, the supercritical regime of 2D and 3D dusty plasmas can separated into a liquid-like and a gas-like phase by a dynamical crossover, the Frenkel line (see cartoon in Fig.~\ref{fig1}).\\

The rest of this paper is organized as follows. In Sec.~\ref{secII}, we briefly introduce our simulation method to mimic 2D and 3D dusty plasmas. In Sec.~\ref{secIII}, we present the proposed supercritical nature of dusty plasmas and its features. We also provide various diagnostics to discriminate the liquid-like from the gas-like states, with the same resulting transition at $20$ times of the melting point for both 2D and 3D dusty plasmas. In Sec.~\ref{secIV}, we provide an interpretation of the agreement between our newly introduced diagnostics and the traditional Frenkel line criteria. In Sec.~\ref{secV}, we give a summary of our findings. In Appendix ~\ref{sec1}, we provide the details of our MD and Langevin simulations. In Appendix~\ref{sec2}, we present the calculated transport results from our simulation data (self-diffusion constant, shear viscosity, and thermal conductivity), as well as a new analysis of previous results from~\cite{Daligault:2012, Donko:2008, Daligault:2014, Donko:2004, Scheiner:2019}.

\section{Simulation method}
\label{secII}
We perform equilibrium molecular dynamics (MD) simulations of 2D and 3D Yukawa liquids to mimic 2D and 3D fluid dusty plasmas as in~\cite{Liu:2005,Goree:2012}. The dynamics of each particle $i$ is governed by the following equation of motion:
\begin{equation}
    m \ddot{\mathbf{r}}_{i}=-\nabla \Sigma_j \phi_{i j}
\end{equation}
where $\phi_{i j}=Q^{2} \,\frac{\exp \left(-r_{i j} / \lambda_{D}\right) }{4 \pi \epsilon_{0} r_{i j}}$ is the Yukawa repulsion between particles $i$ and $j$, separated by a distance $r_{i j}$. Here, $\lambda_{D}$ is the Debye length, $Q$ the charge, and $\epsilon_0$ the vacuum electric permittivity. To characterize dusty plasmas, we use the screening parameter $\kappa=a/\lambda_{D}$ and the coupling parameter $\Gamma = Q^2/ \left(4 \pi \epsilon_0 a k_{B}T \right)$~\cite{Kalman:2004,Merlino:2004,Morfill:2009, Piel:2010, Bonitz:2010}, where $a$ is the Wigner-Seitz radius of $(n \pi )^{-1/2}$~\cite{Kalman:2004} and $(4 n \pi /3)^{-1/3}$~\cite{Goree:2012, Silvestri:2019} as a function of the number density $n$ in 2D and 3D systems, respectively. Clearly, by increasing the screening parameter $\kappa$, the potential changes gradually from a long-range Coulomb-like potential to a hard-sphere-like repulsion. Timescales are normalized using the dusty plasma frequency $\omega_{p d}=\sqrt{Q^2 /\left(2 \pi \epsilon_0 m a^3\right)}$~\cite{Kalman:2004} and $\omega_{p d}=\sqrt{3 Q^2 /\left(4 \pi \epsilon_0 m a^3\right)}$~\cite{Goree:2012} for 2D and 3D dusty plasmas, respectively. For convenience, we will present all our results as a function of the reduced coupling strength $\Gamma/\Gamma_m$, where $\Gamma_m$ corresponds to the melting point determined from the static structure measured from simulations~\cite{Hartmann:2005, Ohta:2000}. In analogy to thermal systems, we can think of the coupling parameter $\Gamma$ as the effective inverse temperature for dusty plasmas, so that a higher $\Gamma$ value corresponds to a lower temperature.

To simulate 2D and 3D dusty plasmas, we confine $N = 4096$ and 8192 particles in two simulation cells with the dimensions of $121.9a \times 105.6a$ and $32.5a \times 32.5a \times 32.5a$, respectively, using periodic boundary conditions. We always set the simulation conditions as melted dusty plasmas, i.e., for each chosen $\kappa$ value between $0.5$ and $3$, we specify various $\Gamma$ values lower than the corresponding melting point $\Gamma_{m}$ of 2D~\cite{Hartmann:2005} and 3D dusty plasmas~\cite{Ohta:2000}. We also use the reduced coupling strength $\Gamma/\Gamma_{m}$~\cite{Donko:2006, Khrapak:2020} to characterize the relative temperature of the studied systems. The integration time step is chosen to be $0.005~\omega_{pd}^{-1}$, small enough as justified in~\cite{Liu:2005}, while the interparticle Yukawa repulsion at the radii beyond $22a$~\cite{Liu:2005} and $8a$ is truncated directly for 2D and 3D systems, respectively. Other simulation details are the same as in~\cite{Liu:2005,Goree:2012}. As the output of our simulations, the obtained time series of positions and velocities for all simulated particles are used to determine various physical quantities reported in the main text. We also perform Langevin dynamical simulations~\cite{Feng:2008} of 2D and 3D dusty plasmas to confirm that our reported results are valid. More details can be found in Appendix \ref{sec1}.

\begin{figure}[h!]
	\centering
	\includegraphics{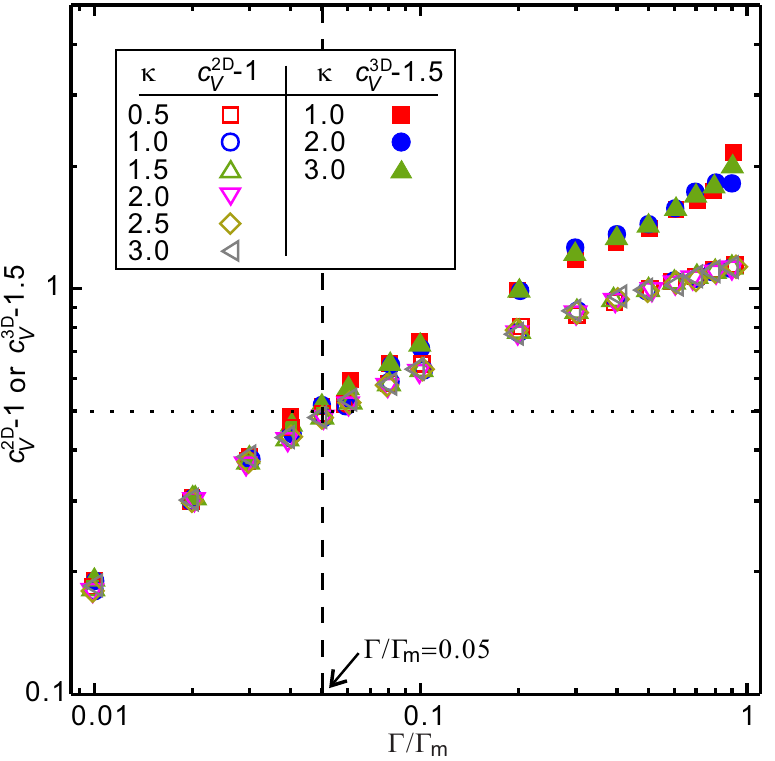}
	\caption{\label{fig2}
Specific heat $c_V$ for 2D and 3D dusty plasmas as a function of the reduced coupling strength $\Gamma/\Gamma_m$ for different values of the screening parameter $\kappa$. The vertical dashed line indicates the critical value $\Gamma/\Gamma_m=0.05$. The horizontal dotted line corresponds to the value $0.5$.}
\end{figure}

\section{Results}
\label{secIII}

\subsection{Heat capacity}
We start our discussion with the analysis of the heat capacity. In what follows, we will refer to the specific heat per particle, so that the total number of particles $N$ disappears from our expressions. Moreover, in dusty plasmas, the temperature is expressed in units of energy, so that $k_B=1$ as well~\cite{Nosenko:2008}. 

At lower temperatures, or equivalently larger coupling parameters, the Frenkel frequency is much lower than the Debye one and collective shear waves display an almost gapless dispersion as in solids. In that regime, for 3D systems, we expect two transverse waves and one longitudinal wave there, each of which contributes $k_B T/2$ potential energy from the equipartition theorem~\cite{Bolmatov:2013,Pathria:2011}, and $k_B T/2$ kinetic energy. For large values of the coupling parameter, $\Gamma/\Gamma_m \gg 1$, the heat capacity approaches therefore the solid-like value, $c_V=3$. When transferring from the rigid liquid to the nonrigid gas-like state, the two transverse waves cannot be sustained anymore, and the heat capacity reduces to $c_V=2$ ~\cite{Brazhkin:2013,Bolmatov:2013}. Qualitatively, this gradual decrease reflects the disappearance of the solid like oscillations into the diffusive gas-like motion. In other words, the number of transverse modes with frequency $\omega>\omega_F$ decreases towards the gas-like regime. As a result, the Frenkel line is determined by the simple condition $c_V^{\rm 3D} = 2$ in 3D systems~\cite{Bolmatov:2013}. Following a similar argument, for 2D systems, there is only one transverse wave, so that $c_V$ decreases from $2$ to $1.5$ when the transverse wave cannot be sustained anymore. Thus, the Frenkel line for 2D systems can be determined from the condition $c_V^{\rm 2D} = 1.5$.\\

In the microcanonical ensemble, the specific heat $c_V$ can be derived from the fluctuations of the kinetic energy (KE) using~\cite{Lebowitz:1967,Cancrini:2017}
\begin{equation} \label{nono}
\frac{2\,( \langle {\rm KE}^2 \rangle - \langle {\rm KE} \rangle ^2 )}{Nd( k_BT ) ^2}=\frac{c_V-d/2}{c_V},
\end{equation}
where $d$ is the dimensionality of the simulated system and $k_B$ the Boltzmann constant. Since the temperature is expressed in units of energy, the obtained specific heat $c_V$ is dimensionless, as in~\cite{Nosenko:2008}. The numerical results from MD simulations are presented in Fig.~\ref{fig2} as a function of the reduced coupling strength $\Gamma/\Gamma_m$, for different values of the screening parameter $\kappa$ and for both 2D and 3D systems. Clearly, as the reduced coupling strength $\Gamma / \Gamma_m$ increases, the specific heat $c_V$ increases monotonically, for both 2D and 3D fluid dusty plasmas. Since the coupling strength has to be interpreted as an inverse temperature, this behavior is consistent with the expected heat capacity for a liquid, which contrary to that of solids, decreases monotonically with temperature. At $\Gamma \approx \Gamma_m$, the obtained values match the expectations from solid state theory, $c_V^{\rm 3D} = 3$ and $c_V^{\rm 2D} = 2$. In the opposite limit, $\Gamma/\Gamma_m \rightarrow 0$, the data approaches the ideal gas result, $c_V^{\rm 3D} = 3/2$ and $c_V^{\rm 3D} = 1$.

From our simulation results in Fig.~\ref{fig2}, $c_V^{\rm 2D} = 1.5$ and $c_V^{\rm 3D} = 2$ both occur at the same reduced coupling strength of $\Gamma / \Gamma_m = \gamma_c=0.05$, with $\gamma_c$ the dimensionless critical value. This clearly indicates that the Frenkel line for both 2D and 3D fluid dusty plasmas is located at the same $\gamma_c$, suggesting that the liquid-like-gas-like transition in dusty plasmas or Yukawa systems is independent of the dimensionality. Intuitively, this could be explained by considering the dynamics of transverse shear waves. In absence of shear excitations, the system is effectively isotropic and therefore insensitive to the number of dimensions. This is further proved by the universal collapse of the two curves in Fig.~\ref{fig2} for $\Gamma / \Gamma_m<\gamma_c$, where no collective shear waves can be sustained any more. On the contrary, in the rigid liquid phase, the heat capacity of the 3D system is consistently larger than the 2D counterpart simply because of the larger number of emerging propagating shear waves.

Intriguingly, all the 3D and 2D curves collapse into a universal one for different values of the screening parameter $\kappa$. This hints towards a possible universality of our results with respect to the particle interaction potential chosen. As we will see, this universality pertains not only the thermodynamic properties but also the transport ones, such as shear viscosity and thermal conductivity.

\begin{figure}[htb]
	\centering
	\includegraphics{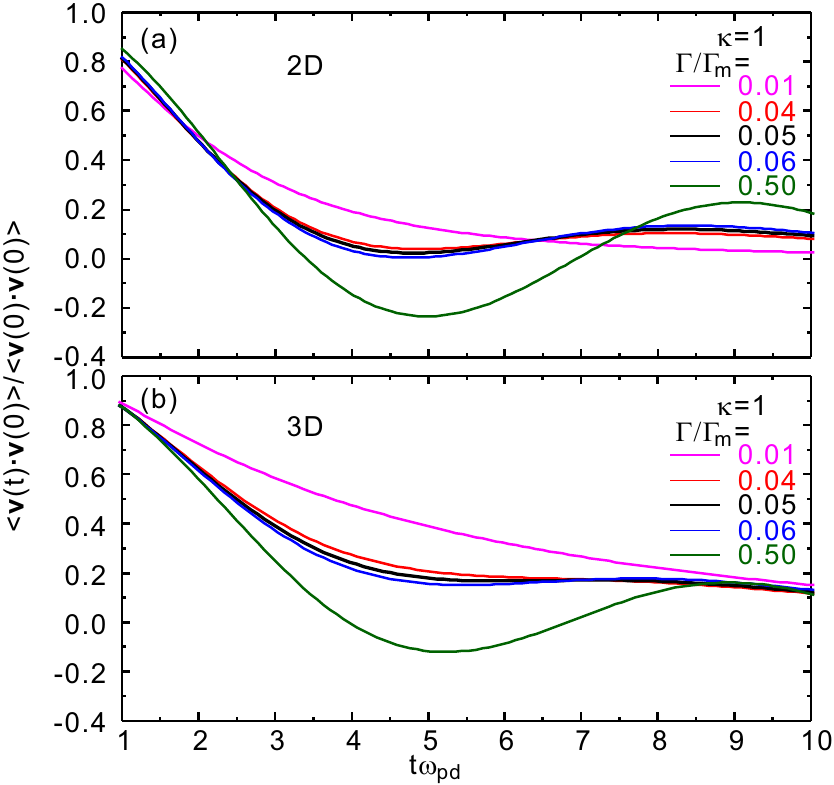}
	\caption{\label{fig3} 
Calculated normalized velocity autocorrelation function (VACF) $C_v(t)$  for 2D (a) and 3D (b) fluid dusty plasmas with $\kappa=1$ and different values of the reduced coupling strength $\Gamma/\Gamma_m$. The black lines correspond to the critical value $\gamma_c=0.05$.}
\end{figure}

\subsection{Velocity autocorrelation function and transport}
In order to verify further our results, we compute the normalized velocity autocorrelation function (VACF):
\begin{equation}\label{def}
    C_v(t)=\frac{\left< {\bf v}\left( t \right) \cdot {\bf v}\left( 0 \right) \right>}{\left< {\bf v}\left( 0\right) \cdot {\bf v}\left( 0 \right) \right>}
\end{equation}
for 2D and 3D fluid dusty plasmas of $\kappa=1$. The numerical results are shown in Fig.~\ref{fig3}.
Our obtained Frenkel line condition, $\gamma_c = 0.05$ from the heat capacity $c_V$ is further confirmed by our results of the VACF in Fig.~\ref{fig3}. In the rigid liquid state, the VACF contains significant oscillations due to the rigidity of the underlying phase. However, in the nonrigid gas-like state, the VACF does not contain oscillations any more, but rather a monotonic decrease typical of gas-like systems. Thus, in~\cite{Brazhkin:2013}, the Frenkel line is proposed to be the critical temperature at which the oscillatory behavior in the VACF just disappears. In~\cite{Brazhkin:2013}, it is also mentioned that, for some cases, such as systems with strong repulsive interactions, this second criterion from the VACF may be not accurate.

In our systems, for 3D fluid dusty plasmas, as $\Gamma/\Gamma_m$ decreases, the oscillations of VACF gradually decay until they completely disappear at $\Gamma / \Gamma_m = 0.05$, perfectly matching the Frenkel line extracted from the heat capacity criterion in Fig.~\ref{fig2}. Besides the results for $\kappa = 1$ shown in Fig.~\ref{fig3}, we have also confirmed that, for all other simulated $\kappa$ values, the critical point at which the oscillations just disappear is always located at $\gamma_c = 0.05$. Thus, the Frenkel line of 3D dusty plasmas determined by the VACF is always at $\gamma_c= 0.05$, in agreement with the analysis of the heat capacity in Fig.~\ref{fig2}. For 2D dusty plasmas, the dynamical transition of the VACF oscillations at $\Gamma / \Gamma_m = 0.05$ is not as distinctive as for the 3D systems. This is consistent with the disclaimers mentioned in~\cite{Brazhkin:2013} and might be attributed to the anomalous diffusion in 2D dusty plasmas~\cite{Ott:2009}, which is absent in the 3D case.

Besides the VACF results presented here, in Appendix \ref{sec2}, we show that the various transport coefficients, such as the diffusion coefficient, the shear viscosity, and the thermal conductivity, calculated from our current simulation data and the previous results of Yukawa systems or one-component plasmas~\cite{Daligault:2012, Daligault:2014, Donko:2008, Scheiner:2019, Donko:2004}, are also optimal diagnostics for the dynamical crossover.

\subsection{Local atomic connectivity and shear relaxation}

It is an important question to ask whether there exists any physical quantity linked to the microscopic dynamics which can identify the Frenkel line crossover. In the Frenkel/Maxwell theories, the re-arrangement time around local minima $\tau$ plays a fundamental role. A more microscopic characterization of such a process is given by the local atomic connectivity $\tau_{LC}$, which is defined as the time for one particle to maintain its surrounding neighbors, or equivalently the time for the atomic topological structure change~\cite{Egami:2013,Ashwin:2015}.

Within the Maxwell viscoelasticity theory, the shear stress relaxation time plays an equivalently important role, specially in the collective dynamics of gapped transverse waves. The shear stress relaxation time $\tau_{M}^{\rm ex}$, first introduced in~\cite{Ashwin:2015} to quantify the viscoelastic response of dusty plasma liquids, controls the relaxation of the excess part (the particle interaction portion) of the shear stress autocorrelation function. Following~\cite{Ashwin:2015}, we calculate the shear stress relaxation time $\tau_{M}^{\rm ex}$ from the ratio of the excess parts of the shear viscosity $\eta^{\rm ex}$ to the infinite frequency shear modulus $G_{\infty}^{\rm ex}$ using
\begin{equation}
    \tau_M^{\rm ex}=\int_0^{\infty} G(t)^{\rm ex} d t/G_{\infty}^{\rm ex}\,,
\end{equation} 
where $G(t)^{\rm ex}$ comes from the autocorrelation function of the particle interaction portion of the shear stress time series~\cite{Ashwin:2015}, while $G_{\infty}^{\rm ex}$ is just the initial value of $G(t)^{\rm ex}$~\cite{Ashwin:2015}. The shear stress relaxation time $\tau_{M}^{\rm ex}$ is approximately equal to the Maxwell relaxation time $\tau_M$ when $\Gamma$ is large, in the so-called potential energy dominated regime~\cite{Ashwin:2015}.

\begin{figure}[htb]
	\centering
	\includegraphics{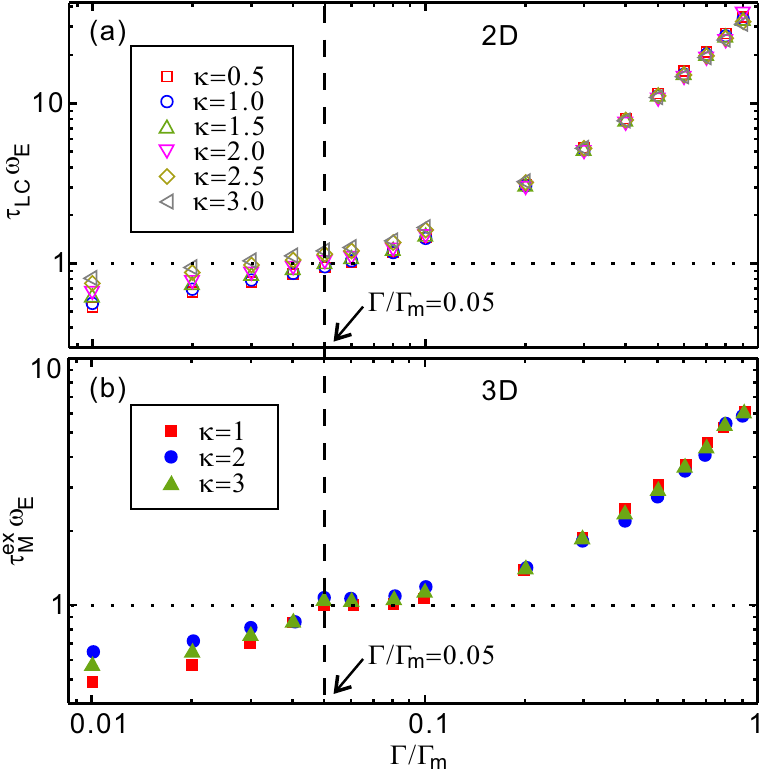}
	\caption{\label{fig4}
Local atomic connectivity time $\tau_{LC}$ and shear stress relaxation time $\tau_{M}^{\rm ex}$ multiplied with the Einstein frequency $\omega_E$ for various conditions of 2D (a) and 3D dusty plasmas (b).}
\end{figure}

From Fig.~\ref{fig4}, we discover that the Frenkel line condition can be expressed as the product of the lifetime of the local atomic connectivity $\tau_{LC}$, or the shear stress relaxation time $\tau_{M}^{\rm ex}$, and the Einstein frequency $\omega_E$ to be unity, i.e., $\tau_{LC} \omega_E = 1$ or $\tau_{M}^{\rm ex} \omega_E = 1$, for 2D or 3D fluid dusty plasmas, respectively. Here, the Einstein frequency $\omega_E$ refers to the oscillation frequency of one particle in the environment where all other particles are assumed to be frozen stationary~\cite{Kalman:2004, Feng:2008}.

As evident from the numerical data shown in Fig.~\ref{fig4}, $\tau_{LC} \omega_E = 1$ in 2D systems and $\tau_{M}^{\rm ex} \omega_E = 1$ in 3D systems both occur at $\Gamma / \Gamma_m = 0.05$, corresponding to the location of Frenkel line determined from $c_V$ in Fig.~\ref{fig2} and VACF in Fig.~\ref{fig3} above, as well as various transport coefficients presented in Appendix \ref{sec2}. Thus, both $\tau_{LC}$ and $\tau_{M}^{\rm ex}$ can be used to quantify the relaxation time $\tau$ between two consecutive hops of a single particle~\cite{Frenkel:1946}. Furthermore, the Einstein frequency $\omega_E$ is just proportional to the inverse of the minimum particle vibration period. As a result, the Frenkel's criterion ~\cite{Frenkel:1946} of $\tau / \tau_D \approx 1$ is qualitatively equivalent to $\tau_{LC} \omega_E \approx 1$ or $\tau_{M}^{\rm ex} \omega_E \approx 1$, consistent with the findings in Fig.~\ref{fig4}. Thus, we propose the conditions $\tau_{LC} \omega_E = 1$ and $\tau_{M}^{\rm ex} \omega_E = 1$ as new diagnostics to determine the Frenkel line for supercritical fluids. To the best of our knowledge, these conditions have not been considered in classical liquids so far.

In Fig.~\ref{fig4}, it is clear that both $\tau_{LC} \omega_E $ in 2D and $\tau_{M}^{\rm ex} \omega_E $ in 3D dusty plasmas exhibit a universal collapse in the rigid liquid-like phase. On the contrary, this universal collapse seems to be less precise in the gas-like regime, $\Gamma/\Gamma_m<0.05$ where the trend of the data depends mildly on the screening parameter $\kappa$. We also notice that the product $\tau_{M}^{\rm ex} \omega_E $ in 3D dusty plasmas displays a discontinuous derivative at the Frenkel line which deserves further investigations.

Finally, we also calculated $\tau_{LC} \omega_E$ in 3D and $\tau_{M}^{\rm ex} \omega_E$ in 2D fluid dusty plasmas, but in that case the universal collapse is less clear. Comparing with the other quantities presented in this work, which all show a universal collapse as a function of $\kappa$, we might speculate that this universality is an emergent property of the collective dynamics which fails, or at least fades away, for microscopic quantities. As a matter of fact, in the gas-like regime, the definition of $\tau_{LC}$ or $\tau_{M}$ is not able to reflect the collective dynamics any more.

\begin{figure}[htb]
	\centering
	\includegraphics{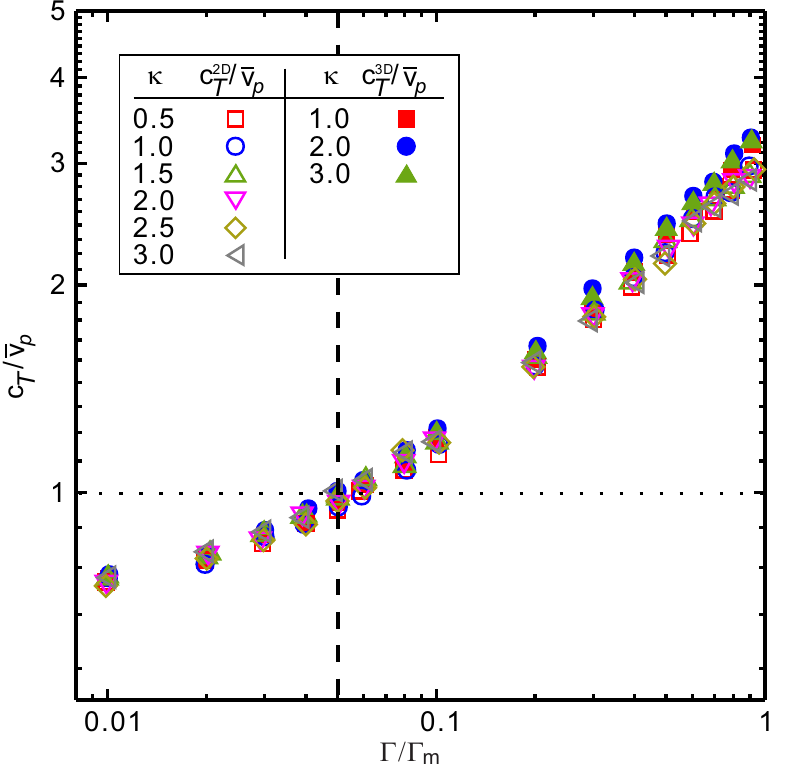}
	\caption{\label{fig5}
	The ratio of the instantaneous transverse sound speed $C_T$ to the average particle speed $\bar{v}_{p}$ for various conditions in 2D and 3D dusty plasmas. The vertical dashed line indicates the location of the Frenkel line as derived from the specific heat and the other diagnostics above.}
\end{figure}

\subsection{Instantaneous transverse sound speed}
Besides the aforementioned conditions $\tau_{LC} \omega_E = 1$ and $\tau_{M}^{\rm ex} \omega_E = 1$, we discover that the unity ratio of the instantaneous transverse sound speed $C_T$ to the average particle speed $\bar{v}_{p}$ can also discriminate the liquid-like and gas-like states and therefore be an optimal diagnostic for the location of the Frenkel line. Here, $C_T$ is the instantaneous speed of transverse sound which can be derived from the infinite frequency shear modulus $G_\infty$~\cite{Egami:2013,Huang:2022,Hansen:1986}. We observe that the unity ratio of $C_T / \bar{v}_{p}$ occurs at the same value of  $\Gamma / \Gamma_m = 0.05$ corresponding to the proposed Frenkel line. Some of these results for 2D dusty plasmas have already appeared in~\cite{Huang:2022}. Here, new results for 3D dusty plasmas, as well as more results of 2D dusty plasmas under different conditions are presented.

When $C_T / \bar{v}_{p} > 1$, the transverse sound speed is faster than the motion of individual particles, corresponding to the ``cooperative dynamics regime''~\cite{Huang:2022}, which is a defining property of rigid liquids. However, when $C_T / \bar{v}_{p} < 1$, in the ``individual dynamics regime''~\cite{Huang:2022} belonging to the nonrigid gas-like state, the average speed of individual particles is larger than the transverse sound speed. Thus, the transition between the cooperative and individual dynamics regimes at $C_T / \bar{v}_{p} = 1$ coincides exactly with the rigid liquid and nonrigid gas-like states separated by the Frenkel line. Our newly proposed criteria for the Frenkel line, $\tau_{LC} \omega_E = 1$, $\tau_{M}^{\rm ex} \omega_E = 1$ and $C_T / \bar{v}_{p} = 1$, may be verified in the future in other physical systems, such as Lennard-Jones and soft-sphere systems~\cite{Brazhkin:2012,Brazhkin:2013}.

As compared with $\tau_{LC} \omega_E$, $\tau_{M}^{\rm ex} \omega_E$, $c_V$, and others, the concept of the speed ratio of collective to individual dynamics used in the diagnostic of $C_T / \bar{v}_{p} =1$ appears to be superior due to at least two reasons. First, from Fig.~\ref{fig5}, the results of $C_T / \bar{v}_{p}$ for 2D and 3D fluid dusty plasmas collapse into one universal ``master curve''. This happens independently of the dimensionality of the system and the value of the screening parameter $\kappa$. Second, the concept of $C_T / \bar{v}_{p}$ can be generalized to different physical processes. For example, during compressional shocks in 2D dusty plasmas, a clear transition at the condition of $v_{left} / C_{l, preshock} = 1$ has been observed in Figs.~4 and 5 of~\cite{Sun:2021}, where $v_{left}$ is the drift velocity of particles after shocks and $C_{l, preshock}$ is the longitudinal wave speed. Clearly, the main cooperative dynamics during compressional shocks~\cite{Sun:2021} is represented by $C_{l, preshock}$, and not $C_T$ as above. At the same time, for compressional related dynamics, the average particle speed should be represented by the drift velocity $v_{left}$ in the postshock region along the shock propagation direction, and not the thermal velocity $\bar v_p$. Thus, in analogy to the $C_T/\bar v_p=1$ criterion, the predicted ``phase'' transition should be located at $C_{l, preshock} / v_{left} = 1$, which is exactly what is observed in~\cite{Sun:2021}. Therefore, when $C_{l, preshock} / v_{left} < 1$, the compressed system exhibits a gas-like behavior in which many particles can penetrate the shock front to enter the preshock region, as confirmed in~\cite{Qiu:2022}. By extrapolation, we do expect that the speed ratio of the cooperative and individual dynamics may be regarded as a universal criterion valid for different physical processes.

\section{Discussion}
\label{secIV}
In this work, MD simulations are performed to investigate the collective and microscopic dynamics in 2D and 3D fluid dusty plasmas. We propose that, below the well-established solid-liquid phase transition (see Fig.~\ref{fig1}), these systems should be regarded as supercritical liquids. To support this hypothesis, we reveal a dynamical transition between the rigid liquid and nonrigid gas-like states in the supercritical regime of dusty plasma, using the concept of Frenkel line. We probe this dynamical crossover with several microscopic and macroscopic thermodynamic and transport quantities such as the heat capacity, the VACF, the shear viscosity, and the thermal conductivity. Furthermore, we propose several new criteria to identify the Frenkel line,
\begin{equation}
    \tau_{LC} \,\omega_E =\tau_{M}^{\rm ex} \,\omega_E=C_T / \bar{v}_{p} = 1 \,,
\end{equation}
providing a new perspective into this old debate. We expect these measures to be generally valid beyond the dusty plasma system considered here.

\begin{figure}[htb]
	\centering
	\includegraphics{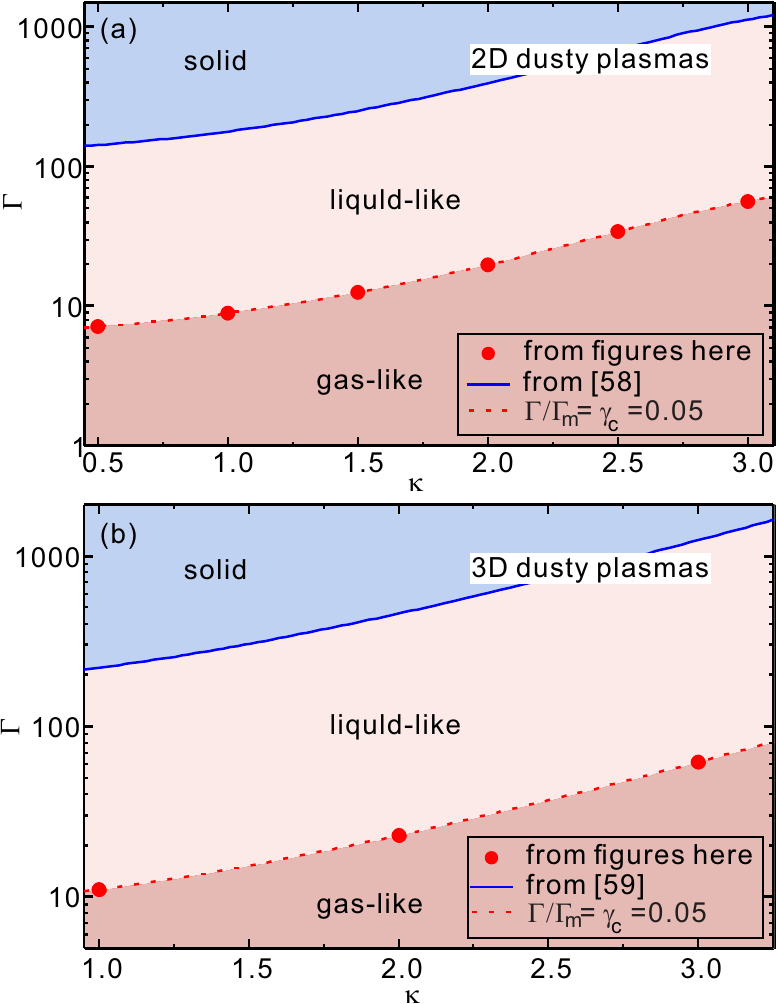}
	\caption{\label{figphase}
Obtained phase diagrams for 2D (a) and 3D (b) dusty plasmas. For both 2D and 3D dusty plasmas, the melting point $\Gamma_m$ for the solid-liquid phase transition (solid blue curves) is obtained from~\cite{Hartmann:2005, Ohta:2000}. The dashed line indicates the location of the dynamical Frenkel crossover, which is universally given by $\Gamma/\Gamma_m=0.05$.}
\end{figure}

Starting from our conjecture, expressed as the cartoon in Fig.~\ref{fig1}, we are now in the position to fundamentally re-define the phase diagram for 2D and 3D dusty plasmas, by adding a new dynamical crossover into it, as shown in Fig.~\ref{figphase}. Importantly, from all our studied diagnostics, it is found that, for both 2D and 3D dusty plasmas, the transition point between the liquid-like and gas-like states is always at $\Gamma/\Gamma_m = 0.05$. This value corresponds to a temperature of $20$ times of the melting point and coincides with the recently proposed criterion to discriminate the strong and weak coupling regimes in dusty plasmas in~\cite{Huang:2022}, identified from the behavior of the shear viscosity. Thus, the proposed weak coupling regime of dusty plasma in~\cite{Huang:2022} is just equivalent to the gas-like state, while the proposed strong coupling regime in~\cite{Huang:2022} corresponds to the liquid-like state. From this point of view, the criterion of $\Gamma / \Gamma_m = 0.05$ in~\cite{Huang:2022} does contain fundamental physical significance to discriminate the strong and weak coupling regimes and it may supersede the traditional criterion $\Gamma = 1$~\cite{Ichimaru:1982}.

The supercritical dynamics for 2D and 3D dusty plasmas revealed in this work presents an emergent degree of universality with respect to the dimensionality of the system and the value of the screening parameter $\kappa$, which hints towards the existence of a general fundamental origin. It would be fruitful to extend this analysis to different systems, potentials and conditions to ascertain, verify, and understand this conjectured universal character.

\section{Summary}
\label{secV}
In summary, we propose a fundamental re-definition of the phase diagram of 2D and 3D dusty plasmas by introducing a new dynamical separation between liquid-like and gas-like phases. Our results provide a strong evidence for the supercritical collective dynamics in strongly coupled plasmas and open the path towards a new interpretation of their fundamental nature which could be fruitful for the modern understanding of plasmas, classical liquids and supercritical phases of matter.\\

\subsection*{Acknowledgments} 
This work was supported by the National Natural Science Foundation of China under Grant Nos. 12175159 and 11875199, the 1000 Youth Talents Plan, startup funds from Soochow University, and the Priority Academic Program Development (PAPD) of Jiangsu Higher Education Institutions. We thank Xiaqing~Shi for helpful discussions. M.B. acknowledges the support of the Shanghai Municipal Science and Technology Major Project (Grant No.2019SHZDZX01) and the sponsorship from the Yangyang Development Fund. M.B. would like to thank Chulalongkorn University for the warm hospitality during the completion of this work.

\bibliographystyle{apsrev4-1}

\newpage
\clearpage
\appendix 
\section{Simulation details}
\label{sec1}

\subsection{MD simulations}

All results presented in our paper are obtained using MD simulations similar to those in~\cite{Liu:2005,Goree:2012,Huang:2022}. For each simulation, first we run $10^{6}$ steps with the Nos\'e-Hoover thermostat, so that the simulated system reaches the equilibrium state under the specified conditions of $\Gamma$ and $\kappa$. Then, the Nos\'e-Hoover thermostat is turned off, and we integrate the next $10^{5}$ steps to confirm the equilibrium state is unchanged. Finally, we integrate the last $10^{6}$ steps, which is the only used for our data analysis reported in this paper. For each simulation run, the output is just the obtained time series of positions and velocities for all simulated particles, which we use to determine the various physical quantities reported in the paper. Note, for our simulation data, when the thermostat is turned off, we always make sure that the system temperature only mildly fluctuates, without any overall drifts.

\subsection{Langevin simulations}

Besides the MD simulations presented in the main text, we also performed Langevin dynamical simulations~\cite{Feng:2008} of 2D and 3D dusty plasmas to confirm that our reported results are accurate. In the Langevin dynamical simulations, the equation of motion~\cite{Feng:2008} for our simulated particles is given by
\begin{equation} 
m\ddot{\mathbf{r}}_{i}=-\nabla \Sigma_j \phi_{i j}-\nu m \dot{\mathbf{r}}_{i}+\xi_{i}(t),
\end{equation}
where the term of $-\nabla \Sigma_j \phi_{i j}$ is the Yukawa repulsion between particles $i$ and $j$, the same as in the MD simulation. The second term $\nu m \dot{\mathbf{r}}_{i}$ is the frictional gas drag, while the last term $\xi_{i}(t)$ is the Langevin random noise obeying the fluctuation-dissipation theorem~\cite{Pathria:2011,Feng:2008}. 

We simulate $N=4096$ and $8192$ dust particles for 2D and 3D systems, confined in the simulation cells with same size and with periodic boundary conditions as in our MD simulations. The specified values of the screening parameter $\kappa $ and the coupling parameter $\Gamma$ in our Langevin simulations are exactly the same as those used in the MD simulations. The time step and the total temporal duration of our Langevin simulation are both chosen to be the same as those for our MD simulation, which are $0.005~\omega_{pd}^{-1}$ and $10^6$ steps, respectively. The only new parameter in the Langevin simulation is the gas damping rate $\nu$, which is chosen as $\nu=0.03~\omega_{pd}$, a typical value in dusty plasma experiments~\cite{Feng:2008}. 

The Langevin simulation data are also used to calculate various physical quantities to confirm our findings in the main text. Different from the microcanonical ensemble of MD simulations, Langevin simulations correspond to the canonical ensemble, where the temperature, or the kinetic energy, of the simulated system is unchanged. As a consequence, Eq.~(2) in the main text cannot be used to determine $c_V$ anymore. Here, we use the fluctuation of the total energy ${\rm E}$ of the simulated system to derive the specific heat $c_V$ as~\cite{Pathria:2011}
\begin{equation} 
c_V = \frac{ \langle {\rm E}^2 \rangle - \langle {\rm E} \rangle ^2 }{N (k_B T)^2}.
\end{equation}
Except for the specific heat $c_V$, all other physical quantities are calculated using the same methods as for our MD simulations. We find that the physical quantities from our Langevin simulations are within numerical uncertainties the same as those reported in this paper, further confirming the reliability of all conclusions in this paper.

\section{Transport properties}
\label{sec2}

Using our simulation data, and revisiting existing data in the literature, we quantify transport properties such as the self-diffusion coefficient, the shear viscosity using the Green-Kubo relation~\cite{Pathria:2011}, and the thermal conductivity. As we will show next, all these quantities are also very efficient diagnostics to discriminate the rigid liquid phase from the gas-like phase in the supercritical regime of 2D and 3D dusty plasmas. 

\subsection{Self-diffusion}

For our 3D simulation data, we calculate the diffusion coefficients $D$ using the corresponding Green-Kubo relation~\cite{Pathria:2011}
\begin{equation}\label{diffusionGK}
D=\frac{1}{3} \int_{0}^{\infty} \left<{\bf v} \left( t \right) \cdot {\bf v}\left( 0 \right) \right> d t.
\end{equation}
In fact, Eq.~(\ref{diffusionGK}) is just the time integral of the velocity autocorrelation function (VACF) $\left<{\bf v} \left( t \right) \cdot {\bf v}\left( 0 \right) \right>$ for the particle motion. We do not study the diffusion coefficient for our 2D simulations. The reason is that as lower dimensional systems, 2D dusty plasmas exhibit anomalous diffusion, as reported in~\cite{Ott:2009}, which could strongly affect our analysis and deserves a separate investigation.

\begin{figure}[htb]
	\centering
	\includegraphics{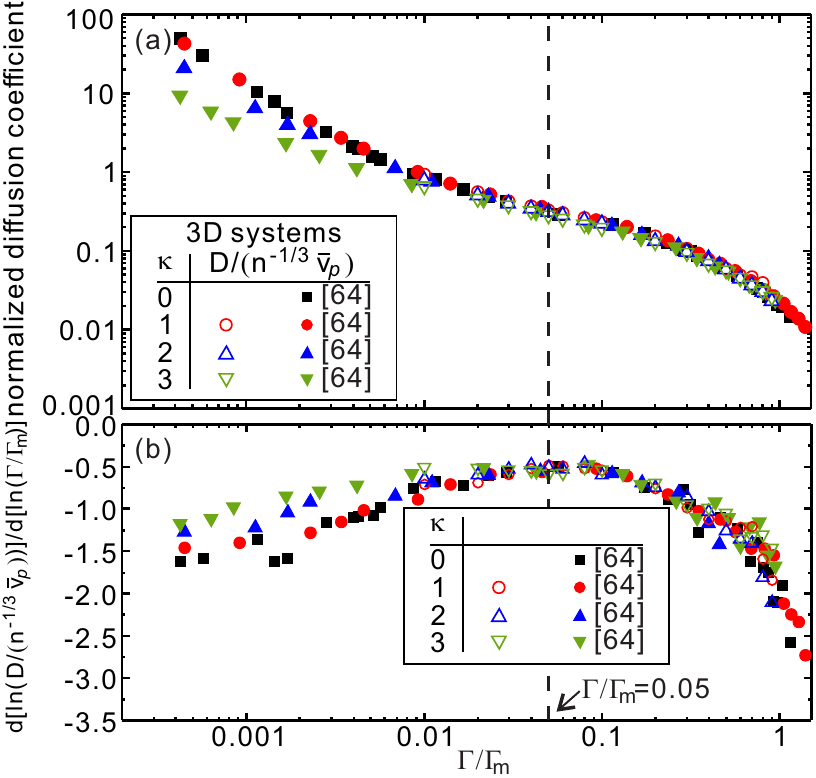}
	\caption{\label{diffusionD}Obtained normalized self-diffusion coefficient $D$ (a) and the corresponding logarithmic derivative $d [ln (D / (n^{-\frac{1}{3}} \bar{v}_{p})) ] / d [ln (\Gamma/\Gamma_m)]$ (b) for our 3D dusty plasma simulations, as well as those performed in~\cite{Daligault:2012}. The dashed line indicates the critical line, $\Gamma/\Gamma_m = 0.05$, as probed by other diagnostics.}
\end{figure}

Our calculated self-diffusion coefficients for 3D dusty plasma simulations when $\kappa = 1$, $2$, and $3$ are presented in Fig.~\ref{diffusionD}(a) as a function of $\Gamma/\Gamma_m$. Following the elementary kinetic theory for a dense medium of particles~\cite{Rosenfeld:2001}, we normalize the obtained self-diffusion coefficient $D$ using $n^{-\frac{1}{3}} \bar{v}_{p}$. Besides our diffusion results, in Fig.~\ref{diffusionD}(a), we also plot the data of 3D Yukawa one-component plasmas reported in~\cite{Daligault:2012}, which well agree with our 3D dusty results. Above the dynamical crossover line, in the rigid liquid phase, $\Gamma/\Gamma_m>0.05$, all data for 3D Yukawa systems, and even for 3D Coulomb system for $\kappa = 0$, collapse into a single universal curve. This universality appears less evident in the gas-like phase in which different curves deviate from each other especially for very small values of $\Gamma/\Gamma_m$. 

The self-diffusion coefficient $D$ clearly exhibits a monotonic decreasing trend as the coupling parameter $\Gamma$ increases. Given that $\Gamma$ has to be interpreted as an inverse effective temperature, this trend is reasonable. When the temperature is lowered, the diffusion is obviously suppressed. More importantly, we observe that the Frenkel dynamical crossover corresponds to a saddle point in the self-diffusion data as a function of $\Gamma$ which locates the condition of lowest decreasing rate. 

To better characterize this important feature, we use the logarithmic derivative $d [ln (D / (n^{-\frac{1}{3}} \bar{v}_{p})) ] / d [ln (\Gamma/\Gamma_m)]$ obtained from the data points in panel (a) of Fig.~\ref{diffusionD}. The resulting data are presented in panel (b) of Fig.~\ref{diffusionD}. We observe a clear maximum in the logarithmic derivative which occurs exactly at $\Gamma/\Gamma_m = 0.05$, further confirming the lowest decreasing rate of the normalized diffusion coefficient $D/ (n^{-\frac{1}{3}} \bar{v}_{p})$ in Fig.~\ref{diffusionD}(a). From both panels in Fig.~\ref{diffusionD}, we confirm that the self-diffusion coefficient represents as well an efficient indicator for the liquid-like to gas-like dynamical crossover in 3D dusty plasmas, validating once more the universal value of $\gamma_c=0.05$. To the best of our knowledge, this criterion has not been discussed yet in the supercritical fluid community. Finally, notice that, using the data of ~\cite{Daligault:2012}, also shown in Fig.~\ref{diffusionD} the supercritical transition at $\Gamma/\Gamma_m = 0.05$ is valid for 3D Coulomb systems as well.

\subsection{Shear viscosity}

Using the Green-Kubo relation, the shear viscosity is obtained from the shear stress fluctuations~\cite{Pathria:2011}. First, we calculate the time series of the shear stress from the velocities and locations of all simulated particles using
\begin{equation}\label{stress}
P_{x y}=\sum_{i=1}^{N}\left[m v_{i x} v_{i y}-\frac{1}{2} \sum_{j \neq i}^{N} \frac{x_{i j} y_{i j}}{r_{i j}} \frac{\partial \phi\left(r_{i j}\right)}{\partial r_{i j}}\right].
\end{equation}
Second, we obtain the autocorrelation function of the shear stress fluctuation $C_s$ using
\begin{equation}
C_{s}(t)=\left\langle P_{x y}(t) P_{x y}(0) \right\rangle.
\end{equation}
Finally, the shear viscosity is derived using the time integral below
\begin{equation}
\eta=\frac{1}{V k_{B} T} \int_{0}^{\infty} C_{s}(t) dt,
\end{equation}
where $V$ is the volume of the simulated 3D system, or the area of the simulated 2D system.

\begin{figure}[htb]
	\centering
	\includegraphics{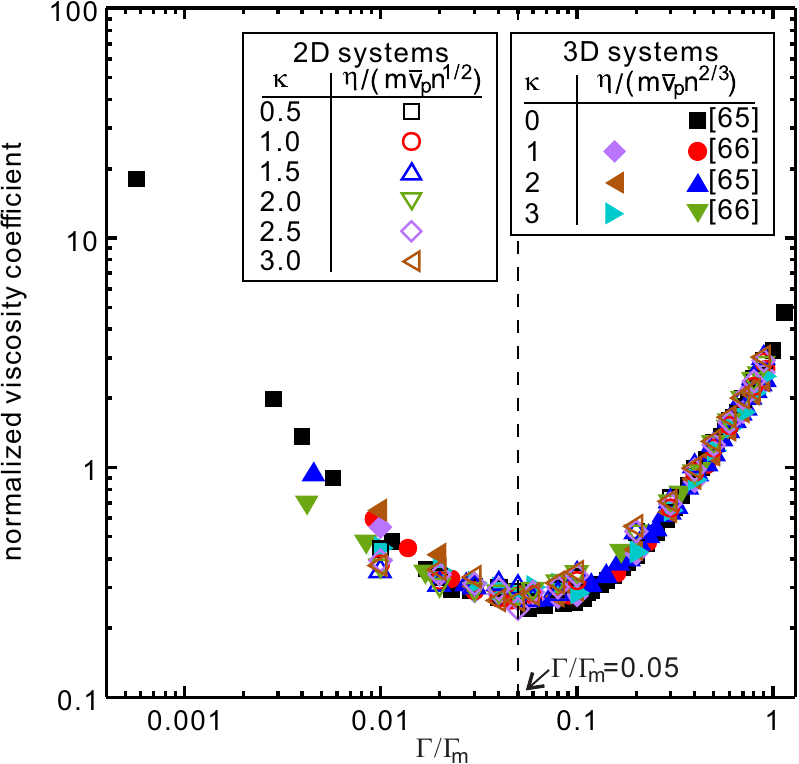}
	\caption{\label{shearvis}The normalized shear viscosity $\eta$ for 2D and 3D dusty plasmas as a function of the coupling parameter $\Gamma$ varies from our MD simulations. The original viscosity results for 3D Yukawa fluids and one component plasmas reported in~\cite{Donko:2008, Daligault:2014} are also plotted here for comparison. The minimum of the normalized shear viscosity always occurs at $\Gamma/\Gamma_m = 0.05$, as shown by the dashed line.}
\end{figure}

Our calculated results for the shear viscosity in 2D and 3D dusty plasmas are presented in Fig.~\ref{shearvis} as a function of $\Gamma/\Gamma_m$ and for different values of $\kappa$. Following ~\cite{Rosenfeld:2001}, we normalize the obtained shear viscosity $\eta$ using $m \bar{v}_{p} n^{\frac{1}{2}}$ and $m \bar{v}_{p} n^{\frac{2}{3}}$ for 2D and 3D dusty plasmas, respectively. Besides our viscosity results, in Fig.~\ref{shearvis}, we also plot the viscosity results for 3D Yukawa fluids and one component plasmas reported in~\cite{Donko:2008, Daligault:2014}, all well agreeing with our results. From Fig.~\ref{shearvis}, all data for both 2D and 3D dusty plasmas, as well as 3D Coulomb systems, collapse into one universal curve in the liquid like phase, $\Gamma/\Gamma_m>0.05$. In the gas-like phase, we observe mild deviations from this universal master curve which deserve further investigations. More importantly, for all conditions in the different systems in 2D and 3D, the minimum of the normalized shear viscosity always occurs at $\Gamma/\Gamma_m = 0.05$, the location of the dynamical liquid-like to gas-like crossover. 

In summary, we find that shear transport is a promising macroscopic quantity to locate the Frenkel line in dusty plasmas as well. In order to rationalize this finding, we resort to classical arguments already presented in \cite{doi:10.1126/sciadv.aba3747}. From the standard kinetic theory, in the gas-like regime, one expects the viscosity to be given by $\eta \sim \rho v_p L$ where $\rho$ is the density, $v_p$ the average particle velocity, and $L$ the mean free path. Since $v_p$ increases with temperature, $v_p \propto \sqrt{T}$, the gas viscosity increases as well. On the contrary, in a liquid phase, the viscosity arises from a quite different underlying dynamics and it decreases with temperature as $\eta \sim \exp \left(U/T\right)$, where $U$ is the activation energy. This immediately implies the existence of a minimum in the viscosity which indicates the crossover between the liquid-like to gas-like behavior. In addition, it turns out that the value of the viscosity at the minimum is universal and given by simple fundamental constants \cite{doi:10.1126/sciadv.aba3747}. It would be interesting to explore further the value of the viscosity at the minimum in strongly coupled plasmas and compare to the results of \cite{doi:10.1126/sciadv.aba3747}. A connection between the minimum of the viscosity in classical liquids and in the strongly-coupled quark-gluon plasma has indeed already been made in \cite{10.21468/SciPostPhys.10.5.118}.

\subsection{Thermal conductivity}
A similar argument to the one just presented for the shear viscosity can be made for energy transport and in particular thermal conductivity/diffusivity \cite{PhysRevB.103.014311}. For this reason, we revisit the thermal conductivity data for 3D Yukawa fluids and 3D one-component plasmas presented respectively in~\cite{Donko:2004} and ~\cite{Scheiner:2019}. In~\cite{Scheiner:2019}, the thermal conductivity is obtained from the fluctuation of the heat flux using the corresponding Green-Kubo relation. In principle, we are also able to determine the thermal conductivity with our simulation data. However, from the suggested time duration needed for the thermal conductivity calculation~\cite{Scheiner:2019}, our simulation run is not long enough, which may lead to large uncertainties. Thus, we only focus on the results reported in~\cite{Donko:2004, Scheiner:2019}. As we will see, those results reveal important physical lessons which were not discussed in the original papers.

\begin{figure}[b]
	\centering
	\includegraphics{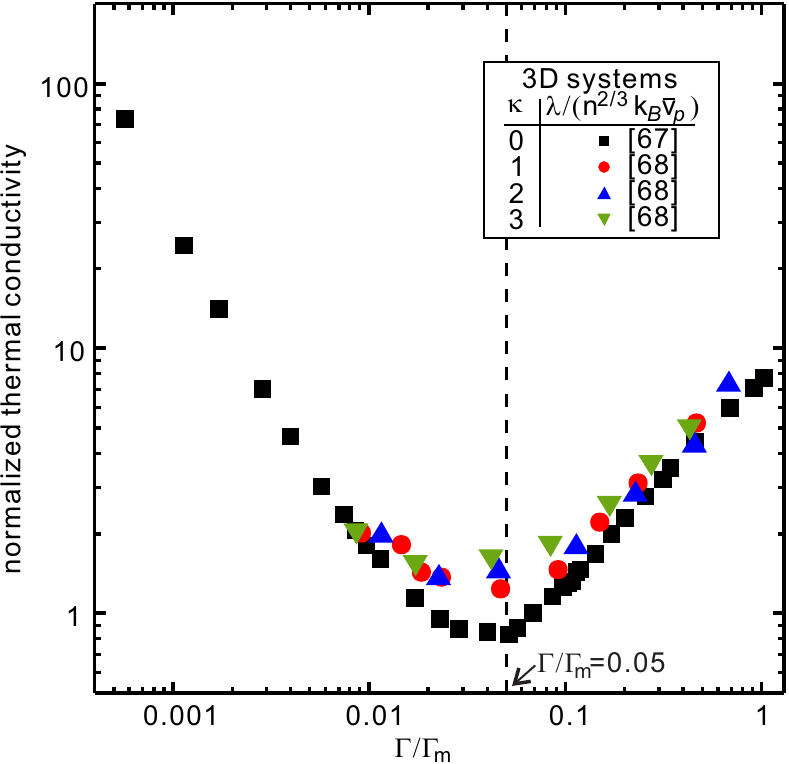}
	\caption{\label{thermalcon}The normalized thermal conductivity for 3D Yukawa fluids and 3D one-component plasmas as the coupling parameter $\Gamma$ varies. The corresponding original data are reported in~\cite{Donko:2004, Scheiner:2019}. The minimum of the normalized thermal conductivity always occurs at $\Gamma/\Gamma_m = 0.05$, as indicated by the dashed line.}
\end{figure}

We conveniently reproduce the data in Fig.~\ref{thermalcon}. Here, we normalize the thermal conductivity $\lambda$ using $n^{\frac{2}{3}} k_B \bar{v}_{p}$~\cite{Rosenfeld:2001}. 

From Fig.~\ref{thermalcon}, the minimum of the normalized thermal conductivity also occurs at $\Gamma/\Gamma_m = 0.05$, which is the same value as the one extracted from the shear viscosity in Fig.~\ref{shearvis} and the diffusion coefficient in Fig.~\ref{diffusionD}. This value also corresponds to the proposed Frenkel line crossover probed with several diagnostics in the main text.

In summary, we find that not only the self-diffusion coefficient and the shear viscosity but also the thermal conductivity displays a distinct feature, i.e., a clear minimum, at the dynamical crossover between liquid-like and gas-like states. Interestingly, the universal collapse observed for other quantities is not so clear in the thermal conductivity data. A more detailed exploration in this direction would be fruitful.


\end{document}